\documentclass{AIAA}
\pdfoutput=1
\usepackage{amsmath}
\usepackage{amsfonts}
\usepackage{graphicx}
\renewcommand{\d}[1]{\ensuremath{\operatorname{d}\!{#1}}}
\graphicspath{{figs/}}
\newcommand{\dario}{\xi}

\begin{document}

\title{Explicit solution to the constant radial acceleration problem}

\author{Dario Izzo\footnote{Scientific Coordinator, Advanced Concepts Team, dario.izzo@esa.int}}
\affiliation{ESA -- Advanced Concepts Team, European Space Research Technology Center (ESTEC), Keplerlaan 1, Postbus 299, 2200 AG Noordwijk, The Netherlands}
\author{Francesco Biscani\footnote{Previously
at the Advanced Concepts Team, ESA. E-mail: bluescarni@gmail.com}}
\affiliation{Schlumberger Abingdon Technology Centre (AbTC), Lambourn Court, Wyndyke Furlong, Abingdon, Oxfordshire, OX14 1UJ, United Kingdom}

\begin{abstract}
While the constant radial acceleration problem is known to be integrable and
has received some recent attention in an orbital mechanics context, a closed form explicit solution,
relating the state variables to a time parameter, has eluded all researchers so far. It is here shown how such a solution exists and is elegantly expressed in terms of the Weierstrass elliptic and related functions.
Previously known facts can be derived from the new explicit solution and
new insights are revealed. 
\end{abstract}

\maketitle

\section*{Nomenclature}
\noindent\begin{tabular}{@{}lcl@{}}
\textit{$e_1,e_2, e_3$}  &=& Roots of the polynomial $f$ \\
\textit{$\tilde e_1, \tilde e_2, \tilde e_3$}  &=&  Roots of the polynomial $g$ \\
\textit{$f(.)$} &=& Third order polynomial associated to the spacecraft dynamics\\
\textit{$g(.)$} &=& Third order polynomial associated to the Weierstrass functions\\
\textit{$g_2, g_3$}  &=& Invariants of the Weierstrass functions \\
\textit{$h$}  &=& Orbital angular momentum \\
\textit{$K(.)$}  &=& Complete elliptic integral of the first type \\
\textit{$r$}  &=&   $|\mathbf r|$\\
\textit{$\mathbf r$}  &=&  Position vector \\
\textit{$t$}  &=& Time \\
\textit{$u,w$}  &=& Auxiliary integration variables \\
\textit{$T_\tau, T_t$}  &=&  Pseudo-period and period for $r$\\
\textit{$v$}  &=&  $|\mathbf v|$ \\
\textit{$\mathbf v$}  &=&  Velocity vector \\
\textit{$w_1,w_2, w_3$}  &=& Roots of the polynomial $g$ in the special case $\gamma_0=0$\\
\textit{$\alpha$}  &=& Constant radial acceleration \\
\textit{$\gamma$}  &=&  Orbital flight path angle \\
\textit{$\theta$}  &=&  Anomaly between $\mathbf r_0$ and $\mathbf r$\\
\textit{$\mu$}  &=& Gravitational parameter \\
\textit{$\rho_1,\rho_2, \rho_3$}  &=& Roots of the polynomial $f$ in the special case $\gamma_0=0$\\
\textit{$\tau$}  &=& Pseudo-time \\
\textit{$\omega, \omega'$}  &=& Periods of the Weierstrass elliptic functions \\
\textit{$\omega_1, \omega_2,\omega_3$}  &=& Periods of the Weierstrass elliptic functions associated to the roots of $g$\\
\textit{$\mathcal E$}  &=&  Specific mechanical energy \\
\textit{$\wp$}, \textit{$\zeta$}, \textit{$\sigma$}  &=& Weierstrass elliptic and related functions \\

\end{tabular} \\

\section{Introduction}

The motion of a point mass particle subject to a central gravity field and to an additional radial acceleration is described
by one of the few known integrable dynamical systems. Its  practical interest is related, among the other things, to spacecraft low-thrust propulsion \cite{quarta2012new,battin1999introduction,prussing1998constant, akella2002anatomy}, to controversial models in modern physics such as that of the Rindler acceleration \cite{carloni2011solar} or anomalies of the gravitational
field in the Solar System such as that of the Pioneer anomaly \cite{nieto2004finding}.

While its solution can be found in terms of Jacobi elliptic integrals,
such a solution is hardly ever discussed nor used as it results in equations expressing the time as a function of the state
variables and not vice-versa. The problem can also be analysed using
basic manipulations of the energy equation \cite{prussing1998constant, akella2002anatomy} which allows to derive classical results \cite{battin1999introduction} and to
discriminate, in special cases, between bounded and unbounded motion. In applications related to spacecraft trajectory design \cite{trask2004optimal,quarta2012new} it is of great importance, on the other hand, to 
have access to an explicit solution to the problem. In the recent work of Quarta and Mengali \cite{quarta2012new} such a solution is proposed in an approximated form making use of circular functions and limited to the bounded case. In that work the authors prefer the use of approximating circular function expressions to the implicit exact solution in terms of Jacobi elliptic functions lamenting the lack of physical insight connected to these mathematical functions. In \cite{san2012bounded}, instead, the solution is computed, again for a special case,
in terms of the Jacobi elliptic functions, confirming how such a solution is implicit and requires the numerical inversion of
complex relations.

In this paper a general explicit solution to the problem is found and discussed. To our knowledge it is the first time such a solution is given.
Our solution is allowed by the careful use of Weierstrass elliptic and related functions 
$\wp, \zeta$ and $\sigma$ (see \cite{whittaker_course_1927} for a good introduction
to these functions). These functions appear in the solution to many problems in classical mechanics and they are a superior tool to express elliptic integrals with respect to the more popular Jacobi expressions \cite{byrd_handbook_1971}, whenever the 3rd or 4th order polynomial expression in the integrand is parametric.  It was recently pointed out in \cite{brizard2009primer}, how
elliptic functions in general and Weierstrass formalism in particular, while part of common knowledge at the beginning of this century, are no longer part of the curricula of engineers or physicists. Hopefully, these results will contribute to spread the use and importance of these beautiful mathematical tools facilitating their use in modern science.

The complete solution to the constant radial acceleration problem is here elegantly expressed by simple explicit equations
describing the complex physical nature of the motion. The newly found expressions are
valid in general for bounded and unbounded motion, they have no restrictive hypothesis and can be
thus used directly in the design of interplanetary trajectories. Interestingly, the solution
to the constant radial acceleration problem involves all the steps needed to solve the more studied Kepler problem:
a) the introduction of an ad-hoc anomaly, b) finding an explicit solution in terms of this anomaly 
and c) the definition of a Kepler's equation to recover the solution in the time domain.

\begin{figure}
\includegraphics[width=\columnwidth]{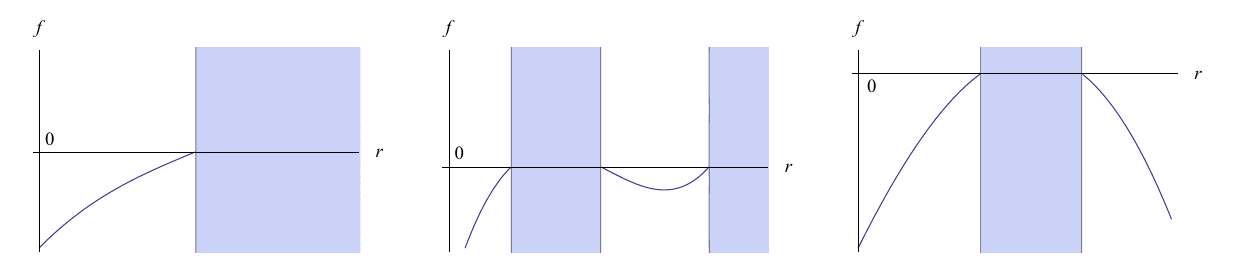}
\caption{Plot of $f(r)$ in the three distinct cases where $f$  has one, two or three positive real roots. Areas are highlighted where $f(r)>0$. \label{fig:sign_f}}
\end{figure}

\section{Problem formulation}
Consider a point mass subject to a Keplerian gravity field and to a constant propulsive acceleration directed radially and indicated with $\alpha$.  Negative $\alpha$ values will account for inward accelerations. Without loss of generality, consider the central field gravitational parameter to be $\mu = 1$. The conservation of the angular momentum $h$ and the conservation of the specific mechanical energy $\mathcal E$ can thus be written as:

\begin{equation}
h = r^2\dot\theta
\label{eq:cons_h}
\end{equation}
\begin{equation}
\mathcal E = \frac{v^2}2 - \frac{1}{r} - \alpha r
\label{eq:vis_viva}
\end{equation}
where the particle distance from the attracting body $r$ is introduced together with
the particle velocity modulus $v$ and the anomaly $\theta$ determining the particle
position with respect to fixed axis. Expressing now $v$ in terms of  $r$ and $\theta$:

\begin{equation}
v^2 = \dot r^2 + r^2\dot\theta^2
\label{eq:velocity}
\end{equation}
substituting Eq.(\ref{eq:velocity}) back into Eq.(\ref{eq:vis_viva}) and expressing $\dot\theta$ in terms of $r$ using Eq.(\ref{eq:cons_h}):
$$
2 \mathcal E = \dot r^2 + h^2 / r^2 - 2 / r - 2\alpha r
$$
and solving for $\dot r$:
\begin{equation}
r\dot r = \operatorname{sgn} ({\dot r}) \sqrt{2\alpha r^3 +2 \mathcal E r^2 + 2 r - h^2} = \operatorname{sgn} ({\dot r}) \sqrt{f(r)}
\label{eq:rdot}
\end{equation}
The solution by quadratures of the constant radial acceleration problem (assuming $r$ monotonically increasing in a given time interval $[t_0,t]$) is:
\begin{equation}
\int_{t_0}^t \d u = \int_{r_0}^r {\frac{u\d u}{\sqrt{2\alpha u^3 +2 \mathcal E u^2 + 2 u - h^2}}} 
\label{eq:quadrature1}
\end{equation}
\begin{equation}
\int_{\theta_0}^{\theta} \d u = h \int_{r_0}^r {\frac{\d u}{u\sqrt{2\alpha u^3 + 2\mathcal E u^2 + 2 u - h^2}}}
\label{eq:quadrature2}
\end{equation}
In the general case in which $r$ is not monotonous in $[t_0, t]$ the integrals above need to be subdivided
accordingly accounting for the sign change. Note that the above integrals define the time as a function of
the state variables, while it is the inverse of such a relation, i.e. expressing the state variables as a function of time, that is of much greater
interest and will here be derived.

\subsection{The polynomial $f(r)$}
Note how the polynomial $f(r)$ defines entirely the point-mass dynamics in the ($r,\dot r$) phase-space via Eq.(\ref{eq:rdot}). A number of interesting properties derive directly from this polynomial.
Indicate with $e_1,e_2$ and $e_3$ the three roots of the third order polynomial $f(r)$
sorted in descending order (first of the imaginary part, then the real part) and with $\Delta$
the discriminant, so that the convention \cite{abramowitz_handbook_1964} reported in Table \ref{tab:econvention} is followed.

\begin{table}
\begin{center}
\begin{tabular}{ccc }\hline
$\Delta > 0$ && $\Delta <0$\\
 & \\
$e_1\ge e_2>e_3$ && $e_1 = a + ib$, \\
&&$e_3 = a - ib$ \\\hline
\end{tabular}
\end{center}
\caption{Convention on the polynomial roots $e_i$ ordering.\label{tab:econvention}}
\end{table}
The three roots of $f(r)$ define entirely the problem taxonomy as
they define the sign of the polynomial $f(r)$: only regions where $f(r) >0$ are allowed. 
In Figure \ref{fig:sign_f}, the three cases that can be encountered are shown:
one, two or three positive real roots for $f(r)$. Only the area $r>0$ and $f(r)>0$ delimits allowed
motion and is shown. For $\Delta < 0$ only one real positive root exists (apply the
Descartes rule and remember that the other two roots must be complex and conjugate) 
and thus the motion is allowed only for $r \ge e_2$. For 
$\Delta > 0$ two cases must be distinguished. The first case is when $\alpha < 0$ (inward acceleration).
In this case, applying again the Decartes rule to the polynomial $f(r)$ defined in Eq.(\ref{eq:rdot}), one can conclude 
that $f$ has always two positive real roots and the
motion is thus bounded as $r \in [e_2,e_1]$. The second case is when $\alpha > 0$ (outward acceleration).
In this case $f(r)$ has three real roots of which either one or three will be positive. 
The motion will then be confined in the area defined by the starting condition  $r_0$. 
It is helpful to visualize the phase-state trajectories plotting Eq. (\ref{eq:rdot}) in a suitable parametrization. 
Write the polynomial $f(r)$ as a function of the initial conditions $r_0$, $v_0$
and the initial flight-path angle $\gamma_0$:
\begin{equation}
f(r) = 2\alpha r^3 + 2(v_0^2/2 - 1/r_0 - \alpha r_0) r^2 + 2 r - v_0^2 r_0^2 \cos^2 \gamma_0
\end{equation}
it is  now possible to plot the resulting trajectory in the phase-space using the radial acceleration $\alpha$
as a parameter and considering the initial conditions as fixed.
\begin{figure}
\includegraphics[width=\columnwidth]{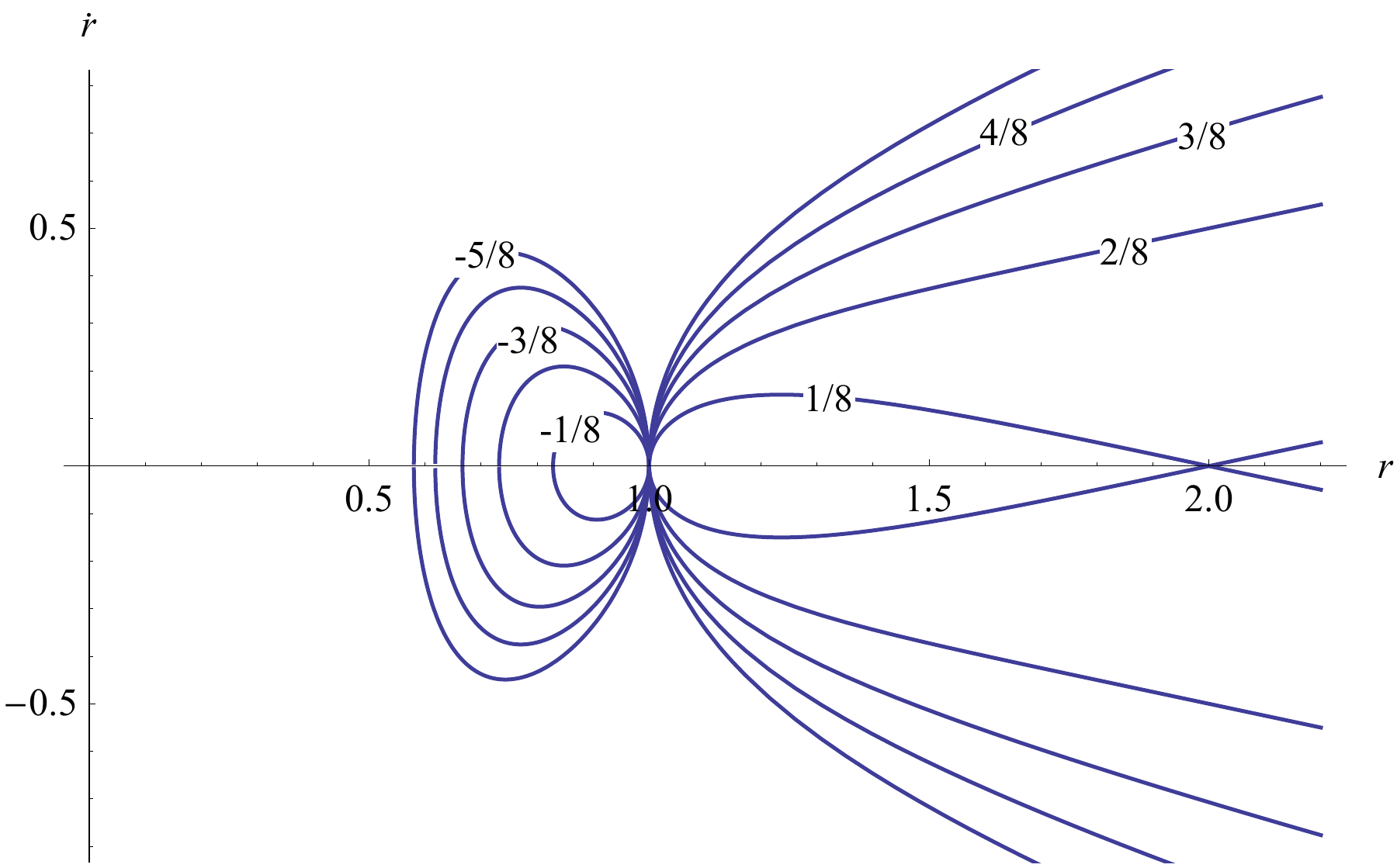}
\caption{Phase-space trajectories for different $\alpha$ values for the case $r_0=v_0=1$, $\gamma_0=0$. The case $\alpha = 1/8$ is also illustrated 
resulting in the homoclinic connection at $r=2$. \label{fig:phase-plot}}
\end{figure}
As an example, such a plot is shown  in Figure \ref{fig:phase-plot} for the particular case treated also by \cite{battin1999introduction, prussing1998constant} where
circular initial conditions are assumed: $r_0v_0^2=1$ and $\cos \gamma_0 = 0$. For this particular case, all trajectories to the left of the line defined by $r=r_0$
correspond to negative values of $\alpha$ resulting in inner orbits, while the half plane $r>r_0$ corresponds the trajectories resulting from a positive $\alpha$ and which eventually
open up and become unbounded. In this case, the three roots of the polynomial $f(r)$ admit a simple expression:

\begin{equation}
\begin{array}{lll}
\rho_1  = r_0, &
\rho_2 = \frac{1- \sqrt{1 -8 \alpha r_0^2}}{4 \alpha r_0}, &
\rho_3 = \frac{1 + \sqrt{1-8 \alpha r_0^2}}{4 \alpha r_0} 
\end{array}
\end{equation}
It is then trivial to conclude that, in order for the resulting motion to be bounded, the following holds:
$$
\alpha r_0^2 < \frac 18
$$
in accordance to the classic result reported for example in \cite{battin1999introduction,prussing1998constant}.

\section{Time as a function of the state (implicit solution)}
Consider now Eq.(\ref{eq:quadrature1}). Following the general integration method
described in \cite{byrd_handbook_1971}, apply the Tschirnaus transformation \cite{cayley_tschirnhausens_1861} to reduce the third degree polynomial to a depressed cubic:
\begin{equation}
u = \sqrt[3]{\frac{2}\alpha} w - \frac{\mathcal E}{3\alpha} 
\end{equation}
and define $\tilde r_0 = \sqrt[3]{\alpha / 2} (r_0 + \frac{\mathcal E}{3\alpha})$ and
$\tilde r = \sqrt[3]{\alpha /2} (r + \frac{\mathcal E}{3\alpha})$. The following then holds:
\begin{equation}
\Delta t = \sqrt[3]{\frac{4}{\alpha^2}} \int_{\tilde r_0}^{\tilde r} {\frac{w\d w}{\sqrt{ 4w^3 - g_2 w - g_3}}}  - \frac{\mathcal E}{3\alpha}\sqrt[3]{\frac{2}{\alpha}} \int_{\tilde r_0}^{\tilde r} {\frac{\d w}{\sqrt{4 w^3 - g_2 w - g_3}}} \label{eq:wkinds}
\end{equation}
where:
\begin{equation}
g_2 = \frac{2}{\alpha} \sqrt[3]{\frac{2}{\alpha}}\left(\frac{\mathcal E^2}{3}-\alpha\right), \hskip1cm
g_3 = h^2 + \frac{2 \mathcal E}{3\alpha} - \frac{4\mathcal E^3}{27\alpha^2}
\end{equation}
Note in Eq.(\ref{eq:wkinds}) the Weierstrass elliptic integrals of the first and second kind \cite{byrd_handbook_1971}. Let's introduce the Weierstrass elliptic function $\wp(z,g_2,g_3)$ with invariants $g_2$ and $g_3$. For a complete treatment of Weierstrass elliptic functions one can refer to \cite{whittaker_course_1927}, it is here sufficient to remember
that  $\wp(z,g_2,g_3)$  is the solution to the following differential equation:  
\begin{equation}
\label{eq:defP}
\wp'^2 = 4\wp^3-g_2\wp-g_3
\end{equation}
and that it is defined in the complex plane where it is holomorphic and doubly periodic
with half-periods indicated with $\omega$ and $\omega'$. In this paper
the notation used by \cite{abramowitz_handbook_1964} is used when dealing with the elliptic functions.
Let us also introduce the Weierstrass $\zeta$ function defined as
$\zeta' (z,g_2,g_3) = - \wp(z,g_2,g_3)$, where the derivative
with respect to the complex variable $z$ is indicated with a prime.
For notation sake, the invariants $g_2$ and $g_3$ will be dropped so that
$\wp(z)$ and $ \zeta(z)$ instead of $\wp(z,g_2,g_3)$ and $\zeta(z,g_2,g_3)$ is used. To solve
the integral in Eq.(\ref{eq:wkinds}), the simple substitution $w = \wp(v)$ and the use of the definition in Eq.(\ref{eq:defP}) leads to:
\begin{equation}
\Delta t = \sqrt[3]{\frac{4}{\alpha^2}} \int_{\rho_0}^{\rho} {\zeta'(v) \d v} -\frac{\mathcal E}{3\alpha}\sqrt[3]{\frac{2}{\alpha}} \int_{\rho_0}^{\rho}  {\d v}
\end{equation}
where $\rho = \wp^{-1}(\tilde r)$ and $\rho_0 = \wp^{-1}(\tilde r_0)$. The inverse of 
the Weierstrass $\wp$ function appears in the above expression indicated with the symbol $\wp^{-1}$.
The problem quadrature is thus found and  can now be formally expressed as:
\begin{equation}
\Delta t =  \sqrt[3]{\frac{4}{\alpha^2}}\left(\zeta(\rho_0)-\zeta(\rho)\right) + \frac{\mathcal E}{3\alpha}\sqrt[3]{\frac{2}{\alpha}}  \left(\rho_0-\rho \right)
\label{eq:time_sol}
\end{equation}
The above expression relates the time as a function of the initial conditions and the current state. It is the
Weierstrassian counterpart to the equivalent expression in terms of the Jacobi elliptic
functions (see \cite{gradshtein_table_2007} \S 3.132 for the general case,
or \cite{san2012bounded} for a particular case) and can be regarded as an 
\lq\lq implicit\rq\rq\ solution to the problem. While more compact than previously known
results (and valid in general for all initial conditions, but singular for $\alpha=0$), to get $r$ as a function of the time $t$ one still needs to invert 
Eq.(\ref{eq:time_sol}) which requires a numerical procedure. 
This problem, shared with the known expressions in terms of the Jacobi elliptic integrals, is the reason the use of 
analytical solutions for the constant radial acceleration problem are not used in practice. Many commented how 
they hinder the physical insight into the problem while not even 
being computationally efficient, thus suggesting the use of approximate approaches.
Contrary to this common knowledge, in the next sections it is shown that it is possible to 
derive the explicit and closed form analytical solution in terms of
the Weierstrass elliptic functions. Such expressions are valid for all
values of $\alpha$ and $\Delta$, for bounded and unbounded motion
and provide, straight-forwardly, a great physical insight into the problem as they explicitly relate the state variables to a pseudo-time (which can also be seen as an orbital anomaly).

\section{The state as a function of a time (explicit solution)}
Consider now Eq.(\ref{eq:rdot}) and introduce the Sundman transformation to regularize the problem:
$$
\d t = r \d\tau
$$
indicating  now with a prime $r'$ the derivative with respect to the new time variable $\tau$:
\begin{equation}
r' = \pm \sqrt{2\alpha r^3 +2 \mathcal E r^2 + 2 r - h^2}
\label{eq:rprime}
\end{equation}
and the quadrature becomes:
\begin{equation}
\int_{\tau_0}^\tau \d u = \int_{r_0}^r {\frac{\d u}{\sqrt{2\alpha u^3 +2 \mathcal E u^2 + 2 u - h^2}}} = \int_{r_0}^r {\frac{\d u}{\sqrt{f(u)}}}
\end{equation}
This integral can be solved and inverted by the direct application of a result which, according to Whittaker \cite{whittaker_course_1927} (p.454, example 2), is due to Weierstrass and which, in our case, may be written:
\begin{align}
\label{eq:weierstrass}
r = & r_0 + \frac 12 \frac 1{\left(\wp(\Delta \tau) - \frac 1{24} f''(r_0)\right)^2} \nonumber \\
& \cdot \left\{ \sqrt{f(r_0)}\wp'(\Delta \tau) + \frac 1{24} f(r_0)f'''(r_0) + \right.\nonumber \\
& \left. + \frac 12 f'(r_0)\left[\wp(\Delta \tau) - \frac 1{24} f''(r_0)\right] \right\}
\end{align}
where $\wp(\tau)$ is the Weierstrass $\wp(\tau,g_2,g_3)$ function with invariants:
\begin{equation}
\begin{array}{l}
g_2 = \frac{\mathcal E^2}{3} - \alpha \\
g_3 = \frac{\alpha^2}{4}(h^2 + \frac{2 \mathcal E}{3\alpha} - \frac{4\mathcal E^3}{27\alpha^2})
\end{array}
\label{eq:gs}
\end{equation}
Define here the polynomial $g(s) = s^3 - g_2 s - g_3$ associated to these invariants and that will be important later in this paper.
Introduce now $r_m$ as the relative minimum for the radius and start counting $t$ and $\tau$ from there. As in a Keplerian orbit $r_m$ would be the pericenter radius, the same name will be used in our case.
By definition, $\dot r(0)=0$ and from Eq.(\ref{eq:rdot}) $f(r_m) = 0$
and the equation above may be written in the simple and elegant form:
\begin{equation}
\label{eq:elegance}
r = r_m + \frac 14 \frac{f'(r_m)}{\wp(\tau) - \frac 1{24} f''(r_m)}
\end{equation}
which expresses one of the state variable (the radius) directly as a function of the Sundman pseudo-time $\tau$.

To search for an equivalent expression for the other state variable $\theta$ let's start  from the momentum conservation:
$$
\frac{\d\theta}{\d t} = \frac h{r^2} \rightarrow \frac{\d\theta}{\d\tau} = \frac hr
$$
\begin{figure}
\centering
\includegraphics[width=0.9\columnwidth]{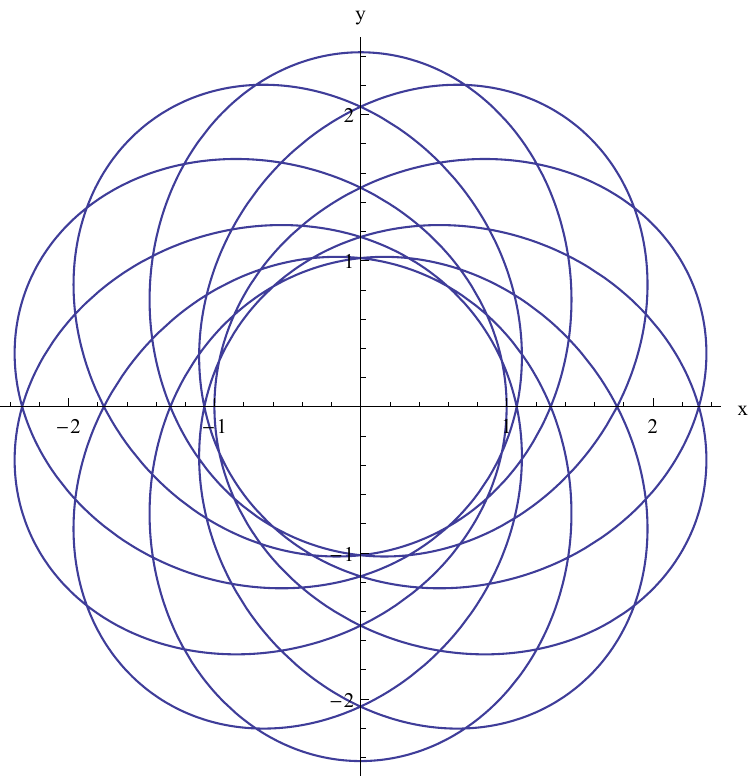}
\caption{Example of an exact periodic solution obtained in terms of the Weierstrass elliptic functions $\wp$, $\sigma$ and $\zeta$.
Initial conditions are $r_m = 1.0$, $v_m = 1.26014$ $\alpha = -.05$. \label{fig:traj}}
\end{figure}
Using Eq.(\ref{eq:elegance}) it is easy to see that:
$$
\int \frac 1r \d\tau = \int \frac{\wp(\tau) + \beta}{\lambda\wp(\tau) + \delta} \d\tau
$$
which is a known integral (see \cite{gradshtein_table_2007} \S 5.141), and hence obtain the analytical expression:
\begin{equation}
\theta = h\frac \tau\lambda + h \frac{\beta\lambda-\delta}{\lambda^2\wp^{\prime}(\dario) } 
\left[\ln\frac{\sigma(\dario-\tau)}{\sigma(\tau+\dario)} + 2\tau\zeta(\dario)\right]
\label{eq:theta_dyn_simple}
\end{equation}
where:
\begin{align}
\beta & = -\frac{1}{24}f^{\prime\prime}\left(r_m\right), & \lambda& = r_m,\label{eq:aux1}\\
\delta & = f^\prime\left(r_m\right) / 4+\beta r_m,& \wp(\dario)&=-\frac{\delta}{\lambda},
\label{eq:aux2}
\end{align}
and $\sigma$ is the Weierstrass $\sigma$ function defined as $\sigma^{\prime} / \sigma = \zeta$. 
Note that $\theta_m=0$ is also assumed, as $\theta$ is counted from the pericenter.  Eq.(\ref{eq:theta_dyn_simple})
above could already be considered the solution as it relates simply and with one short expression the state variable $\theta$ to the pseudo time $\tau$. 
The presence of the logarithm of a complex variable makes the expression not \lq\lq usable\rq\rq\ as the phase ambiguity deriving from the use of a complex logartihm cannot be resolved. The following
few steps address this issue. From the definition of the Weierstrass elliptic function ${\wp^{\prime}}^2(\dario) = 4 \wp^3(\dario) - g_2 \wp(\dario) - g_3$ substituting relevant quantities from Eq.(\ref{eq:gs}) and Eqq.(\ref{eq:aux1})-(\ref{eq:aux2}), the following holds
$$
{\wp^{\prime}}(\dario) = i\frac{v_m (\alpha r_m^2 + r_m v_m^2-1)}{2 r_m}
$$
where $v_m$ is the pericenter velocity. Using this expression and Eqq.(\ref{eq:aux1})-(\ref{eq:aux2}) eventually, the following remarkable identity is found:
$$
h \frac{\beta\gamma-\delta}{\lambda^2\wp^{\prime}(\dario) } = \pm i
$$
To select among the two possible values of $\dario$  in the fundamental rectangle such that  
$\wp(\dario) = -\delta / \lambda$, select in the above expression the plus sign.
The identity above allows to rewrite Eq.(\ref{eq:theta_dyn_simple}) in the form:
\begin{equation}
\label{eq:thetaexp}
\exp{i(v_m\tau-\theta)} = \frac{\sigma(\dario-\tau)}{\sigma(\tau+\dario)} \exp{2\tau\zeta(\dario)} 
\end{equation}
which is not affected by any phase ambiguity any more.
Thus, the solution to the constant radial acceleration  problem, in the new pseudo-time, is described in its most general case
by the following compact expressions:
\begin{equation}
\label{eq:analytical_solution}
\left\{
\begin{array}{l}
r = r_m + \frac 14 \frac{f'(r_m)}{\wp(\tau) - \frac 1{24} f''(r_m)} \\
\exp{i(v_m\tau-\theta)} = \frac{\sigma(\dario-\tau)}{\sigma(\tau+\dario)} \exp{2\tau\zeta(\dario)} 
\end{array}
\right.
\end{equation}
One may explicit further the second of the above relations by the use of the Euler formula for the exponential and obtain:
\begin{equation}
\label{eq:theta_sol}
\begin{array}{l}
\sin\theta =z_{R}(\tau)\sin (v_m\tau) - z_{I}(\tau) \cos (v_m\tau)\\
\cos\theta = z_{I}(\tau)\sin(v_m\tau) +z_{R}(\tau)\cos(v_m\tau)
\end{array}
\end{equation}
having introduced $z_{R}$ and $z_{I}$ as, respectively, the real and the imaginary part
 of $z(\tau)= \frac{\sigma(\dario-\tau)}{\sigma(\tau+\dario)} \exp{2\tau\zeta(\dario)}$. 
In Figure \ref{fig:traj}, an example of a trajectory plotted using the new expressions found is shown.
It is worth to mention here again that Eq.(\ref{eq:analytical_solution}) is \lq\lq universal\rq\rq\ in the sense that
it is valid for bounded and unbounded motion.

\section{The radial Kepler's equation}
As shown in the previous sections, the constant radial acceleration problem admits explicit
solutions relating the state variables $r$ and $\theta$ directly to the pseudo-time $\tau$.
One may look at $\tau$ as the eccentric/hyperbolic anomaly of the Keplerian problem: also in that
case it is a Sundman transformation that relates these anomalies
to the time $t$ (see Berry and Healy \cite{berry2002generalized} for a generic discussion on the relation between orbital anomalies and Sundman transformations). In particular, in the case of the eccentric anomaly, such a relation is  $\d t = \frac r {na}\d E$ where $n$ is the mean motion and $a$ the orbital semi-major axis. In this Keplerian case, recovering the time from the eccentric anomaly,  Kepler's equation needs to be solved. In the case of the constant radial acceleration problem
things are rather similar: an equivalent to Kepler's equation relates the pseudo-time $\tau$ to the time $t$. This equation will be referred to as the \lq\lq radial Kepler equation.\rq\rq\ Using the newly found expression in Eq.(\ref{eq:elegance}), the following can be derived:
$$
t(\tau) = \int_{0}^\tau  \left[r_m + \frac{1}{4}\frac{f^{\prime}(r_m)}{\wp(u) - \frac 1{24} f^{\prime\prime}(r_m)}\right] \d u
$$
It is possible to prove, by direct substitution, that  $\frac 1{24} f^{\prime\prime}(r_m)$ is always a root of the polynomial $g(s)$, say $\tilde e_k$. More specifically, following to the convention
in Table \ref{tab:econvention}, such a root will be $\tilde e_3$  if the motion is unbounded, $\tilde e_2$  if the motion is bounded. It follows that the above integral may be written in the form:
$$
t(\tau) = \int_{0}^\tau  \left[r_m + \frac{1}{4}\frac{f^{\prime}(r_m)}{\wp(u) - \tilde e_k}\right] \d u
$$
This integral is known (see \cite{byrd_handbook_1971} \S 1037.07-09). Exploiting the identity $\tilde e_i \tilde e_j = \frac{g_3}{4 \tilde e_k}$ the expressions reported in \cite{byrd_handbook_1971} are further simplified:
$$
t(\tau) = r_m \tau  - \frac{\tilde e_k f^{\prime}(r_m)}{g_3+ 8 \tilde e_k^3}\left[ \tilde e_k\tau+\zeta(\tau)+\frac 12\frac{\wp^\prime(\tau)}{\wp(\tau)-\tilde e_k}\right]
$$
where $\tilde e_k = \frac 1{24} f^{\prime\prime}(r_m)$. The above equation is undetermined in $\tau =0$ (and thus numerically unstable) 
as both $\wp$ and its derivative $\wp'$ are infinite. To remove this problem  use the identity:
\begin{equation}
\frac{\wp^{\prime}(a)}{\wp(b) - \wp(a)} = \zeta(b-a) - \zeta(b+a) + 2 \zeta(a) = 2 \zeta(a) + \frac{ \mbox{d} }{\d\tau}\ln \left(\frac{\sigma(b-a)}{\sigma(b+a)} \right)
\end{equation}
and $\tilde e_k = \wp(\omega_k)$ (see \cite{abramowitz_handbook_1964} \S18.3.1) to conclude:
\begin{equation}
t(\tau) = r_m \tau  - \frac{\tilde e_k f^{\prime}(r_m)}{2g_3+ 16 \tilde e_k^3}\left[ 2 \tilde e_k\tau +
\zeta(\tau-\omega_k) + \zeta(\tau + \omega_k) \right]
\label{eq:radial_kepler}
\end{equation}
which now holds the correct value $t(0) = 0$. For the sake of the reader's convenience, the definitions of $\omega_k$ (from \cite{abramowitz_handbook_1964} Figure 18.1) as a function of the complex
half-periods $\omega$ and $\omega'$ of the elliptic function $\wp$ are reported:
\begin{equation}
\begin{array}{l}
\omega_1 = \omega \\
\omega_2 = \omega + \omega'\\
\omega_3 = \omega'
\end{array}
\end{equation}
Remarkably, Eq.(\ref{eq:radial_kepler}) is \lq\lq universal\rq\rq\ being formally valid in this form for bounded and unbounded motion. It is Eq.(\ref{eq:radial_kepler}) that is here called \lq\lq the radial Kepler equation\rq\rq. The role it plays in the solution
of the constant radial acceleration problem is the same as that of the Kepler equation in the Kepler problem.

\begin{figure}
\includegraphics[width=0.9 \columnwidth]{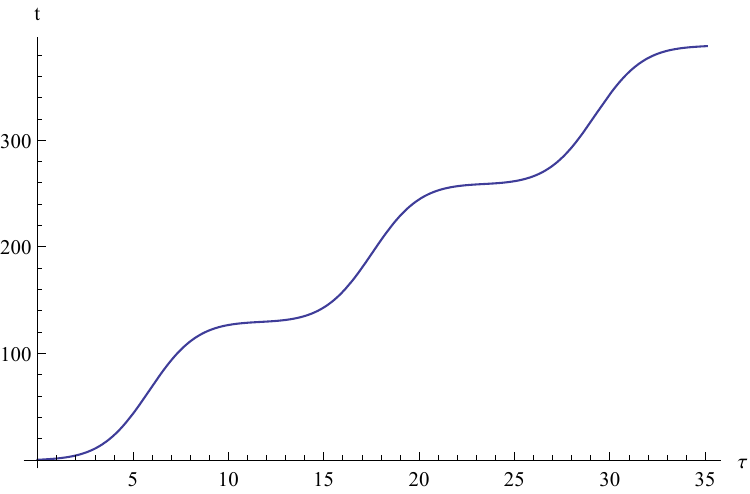}
\caption{A plot of the radial Kepler equation Eq.(\ref{eq:radial_kepler}) in the case $r_m=1.0$; $v_m = 1.56$; $\alpha = -0.01$ }
\end{figure}

\section{Use of the new solution}
\subsection{Periodicity of $r$}
In case of bounded motion it is of interest to compute the period of $r$. In the $\tau$ domain $r$ is
periodic and its period is the real period of $\wp(\tau,g_2,g_3)$ as can be derived trivially from Eq.(\ref{eq:elegance}). 
As the periodicity of $\wp$ for the case of bounded motion is analysed, our analysis is restricted to the case
of a positive discriminant for $g(s)$. 
Compute the two half-periods $\omega$ and $\omega'$ 
of the doubly periodic complex function $\wp(z,g_2,g_3)$ using the known relations with
the complete elliptic integral of the first type $K$ valid for $g_3>0$:
\begin{equation}
\begin{array}{l}
\omega = K(m) / \sqrt{\tilde e_1 - \tilde e_3} \\
\omega' =  i K(1-m) / \sqrt{\tilde e_1 - \tilde e_3}
\end{array}
\end{equation}
where $m = (\tilde e_2-\tilde e_3)/(\tilde e_1 - \tilde e_3)$. The period on the real axis will then be $T=\omega$. In the case $g_3<0$, use the homogeneity condition $\wp(z,g_2,g_3) = -\wp(iz,g_2,-g_3)$. Eventually it is shown that in all cases (i.e. $\forall g_3$), the following holds:
\begin{equation}
\label{eq:pperiod}
T _\tau= 2 K(m) / \sqrt{\tilde e_1 - \tilde e_3}
\end{equation}

\begin{figure}
\centering
\includegraphics[width=0.9\columnwidth]{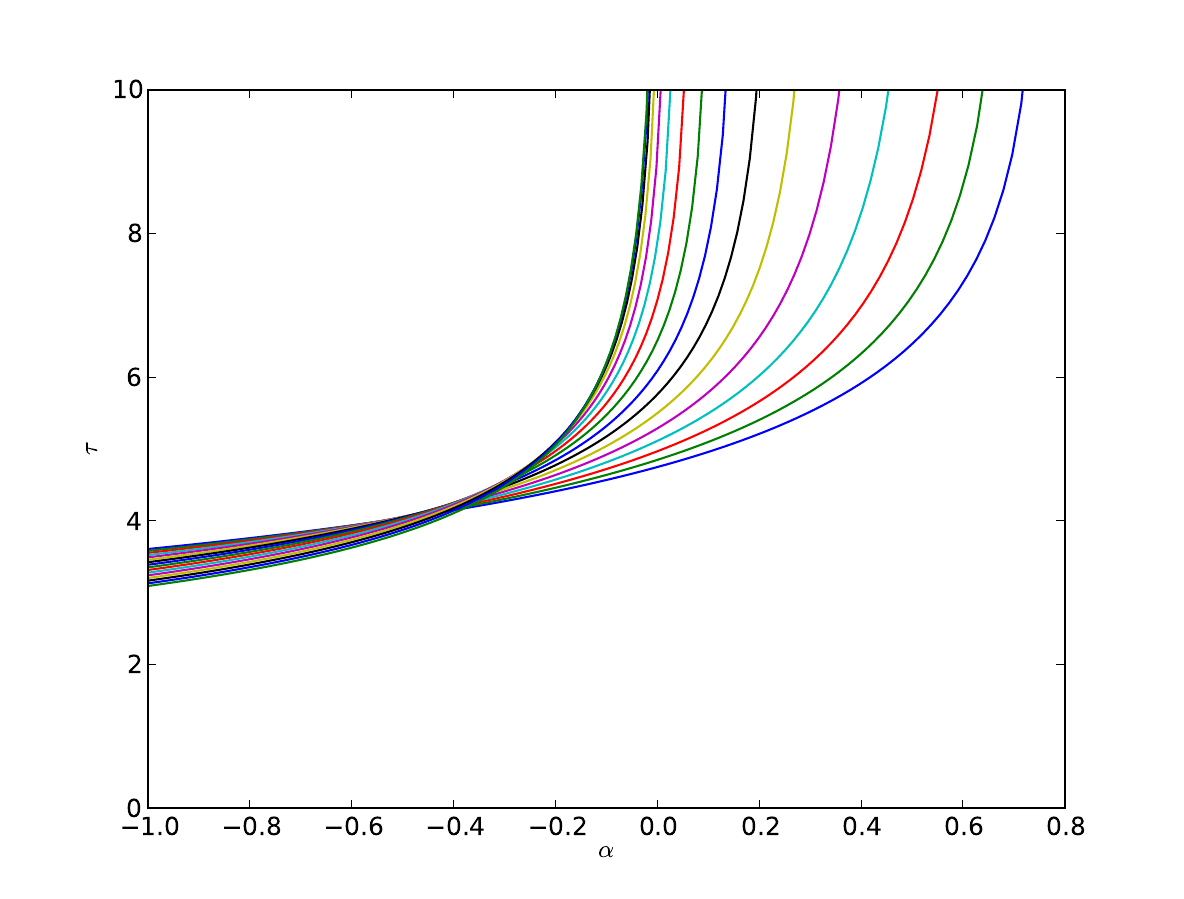}
\caption{Pseudo-period of $r$ at different thrust levels (Eq.(\ref{eq:pperiod}) for different $\alpha$). Different curves corresponds to different starting $v_p$
sampled in $[0.5,1.5]$. \label{fig:period}}
\end{figure}
\noindent
In Figure \ref{fig:period} the pseudo-period $T_\tau$ are plotted against the value
of the radial acceleration $\alpha$ for different initial conditions. For any chosen
value of $\tau$  a value of $\alpha$ always exists resulting in an orbit with that period. Computing then the radial Kepler
equation for $\tau = T_\tau$, an expression for the true period $T_t$ is found:
$$
T_t = r_m T_\tau -  \frac{\tilde e_k f^{\prime}(r_m)}{2g_3+ 16 \tilde e_k^3}\left[ 2 \tilde e_k T_\tau +
4 \zeta(T_\tau / 2) \right]
$$
where the quasi-periodicity of the $\zeta$ functions is exploited (see \cite{abramowitz_handbook_1964} \S 18.2.19). A different, but numerically equivalent, expression
for $T_t$ can be also found computing from Eq.(\ref{eq:time_sol}) the time to travel from pericenter to apocenter (i.e. half-period):
$$
T_t / 2 = \sqrt[3]{\frac{4}{\alpha^2}}\left(\zeta(\rho_m)-\zeta(\rho_M)\right) + \frac{\mathcal E}{3\alpha}\sqrt[3]{\frac{2}{\alpha}}  \left(\rho_m-\rho_M \right)
$$
where the subscript $m$ is used to denote quantities at the closest apporach radius $r_m$.
\subsection{Computing $r_m$, $v_m$ and $\tau_0$}
In  Eq.(\ref{eq:elegance})-(\ref{eq:aux}) the radius at the closest approach (or pericenter radius, indicated with the symbol $r_m$) appear. It can be determined by looking at the roots $e_i$ of the
polynomial $f(r)$ and setting $r_m = e_i$ where $e_i$ is the closest real root to $r_0$
such that $e_i \le r_0$. The initial conditions will, in general, be not given at the pericenter, in which case the initial pseudo-time can be computed directly from Eq.(\ref{eq:elegance}) as:
\begin{equation}
\wp(\tau_0) = \frac 1{24}f^{\prime\prime}(r_m) + \frac 14 \frac{f^\prime(r_m)}{r_0-r_m}
\label{eq:taum}
\end{equation}
the appropriate value for the inversion of $\wp^{-1}$ is selceted by looking only within the first $T_\tau$ and choosing the solution with the correct $\dot r$.

\subsection{The condition for bounded motion}

\begin{figure}
\includegraphics[width=\columnwidth]{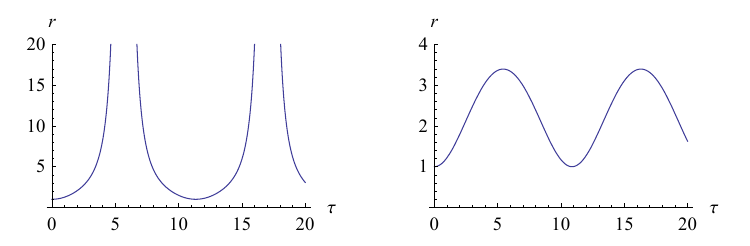}
\caption{Plot of Eq.(\ref{eq:elegance}) for the case $r_0=r_m=1$, $v_0=1.2$. The unbounded case (left) corresponds to $\alpha=0.1$ and 
the bounded case (right) corresponds to $\alpha = 0.02$ \label{fig:r_plots}}
\end{figure}

Once $r_m$ is computed the pseudo-time dependency of $r$ can be computed. In Figure \ref{fig:r_plots},  Eq.(\ref{eq:elegance}) is plotted assuming as a starting position the pericenter radius $r_0 = r_m$,
and as a starting velocity $v_0=1.2$. Two cases are shown: one unbounded, obtained for $\alpha = 0.1$ and one bounded, obtained for $\alpha = 0.02$.
In order for the motion to be bounded it is clear from Eq.(\ref{eq:weierstrass}) that the denominator cannot vanish. Introducing $\wp_{min}$ as the minimum value assumed by $\wp$ on the real
axis, the condition to have bounded motion can be written as $24 \wp_{min} > f''(r_0) = 12 \alpha r_0 + 4 \mathcal E$.
The minimum value assumed on the real axis by the Weiestrass elliptic function is computed introducing the three roots $\tilde e_i$ of the polynomial
$g(s) = 4s^3 - g_2 s - g_3$. The greatest real root is $\wp_{min}$: indicate it with $\tilde e$ and a simple final relation is derived:
\begin{equation}
\label{eq:bounded}
\tilde e > \frac 12 \alpha r_0 + \frac 16 \mathcal E = \frac 1{24} f^{\prime\prime}(r_m)
\end{equation}
which is the generic condition to obtain bounded motion in the constant radial acceleration problem. 
Note how an equivalent to this relation was previously known only for the special case of a starting circular orbit. That result is now
extended to the most general case thanks to the use of Weierstrass elliptic functions.
Take as an example $\alpha = 0.02$, $r0 = 1.1$, $r_p = 1$, $v_p = 1.2$. Evaluating the three roots of $g(s)$:
$$
\tilde e_1 = -0.0402894, \mbox{ }
\tilde e_2 =-0.0170428, \mbox{ }
\tilde e_3 = 0.0573322 \\
$$
and thus $\tilde e  =0.0573322$. Compute now $\frac 12 \alpha r_0 + \frac 16 \mathcal E = -0.039333$ to immediately conclude
that the motion will be bounded by direct application of Eq.(\ref{eq:bounded}). The search for a particular value of $\alpha$ or of the initial velocity $v_0$
which results in an escape trajectory can then be made efficiently, e.g. using a simple bisection algorithm. 

Consider the more restrictive case in which $r_0 = r_m$ and $v_0 = m$.
The three roots of $g(s)$ may be expressed in a simple form, by exploiting the relation $h = r_0 v_0$:
\begin{equation}
\begin{array}{l}
w_1 = \frac 12 \alpha r_0 + \frac 16 \mathcal E = \frac 1{24} f^{\prime\prime}(r_m) \\
w_{2,3} = -\frac 12 (\frac 12 \alpha r_0 + \frac 16 \mathcal E) \pm
 \frac{1}{8 r_0} \sqrt{(2 - r_0 v_0^2)^2 -8 \alpha r_0^3 v_0^2}
\label{eq:roots}
\end{array}
\end{equation}
Note how $\frac 1{24} f^{\prime\prime}(r_m)$ is always a root of $g(s)$, a fact that will have a great importance later.
Applying again Eq.(\ref{eq:bounded}) the motion is proved to be unbounded if and only if $\tilde e = w_1$.
This last condition, after some manipulations, can be shown to be 
equivalent to the set of conditions:
\begin{equation}
\begin{array}{lcl}
r_0v_0^2 < \frac 23 & \mbox{and} &  \alpha < \min{\left(\frac{1-r_0v_0^2}{r_0^2}, \frac{(2 - r_0 v_0^2)^2}{8 r_0^3 v_0^2} \right)}\\
\frac 23 \le r_0v_0^2 \le 2 & \mbox{and} &  \alpha < \frac{(2 - r_p v_0^2)^2}{8 r_0^3 v_0^2} \\
r_0v_0^2 > 2 & \mbox{and} &  \alpha < 0 \\
\end{array}
\end{equation}
In case of a starting circular orbit, we have $r_0v _0^2 = 1$ and the above conditions all collapse into the classical result $\alpha r_0^2 < \frac 18$.
The classical result derived in \cite{battin1999introduction, prussing1998constant} is thus generalized.

\subsection{The condition for periodic motion}

While, in a bounded motion case, $r$ is always a periodic function of both the time and the pseudo-time, the whole trajectory will only be periodic if and only if there exist two numbers 
$M, N \in \mathbb N$ such that $\theta(N T_\tau) = 2 M \pi$. Let us compute the value $\Delta \theta$ reached by the variable $\theta$ after $N$ full periods $T_\tau$. Starting from Eq.(\ref{eq:thetaexp}):
$$
e^{i(v_mNT_\tau-\Delta\theta)} = \frac{\sigma(\dario-NT_\tau)}{\sigma(NT_\tau+\dario)} e^{2NT_\tau\zeta(\dario)} 
$$
Consider:
$$
\phi(\tau) = \frac{\sigma(\dario - \tau)}{\sigma(\dario + \tau)} 
$$
first  compute $\phi(\tau + T_\tau)$ using the quasi-periodicity of the $\sigma$ function (see \cite{abramowitz_handbook_1964} \S 18.2.20) and
the fact that, for $g_3 > 0$, $N T_\tau = 2 N \omega$:
\begin{equation*}
\phi(\tau+N T_\tau) = \phi(\tau+2 N \omega) = - \frac{\sigma(\tau - \dario + 2 N \omega)}{\sigma(\tau + \dario + 2 N \omega)} 
= - \frac{\sigma(\tau - \dario)}{\sigma(\tau + \dario)} \frac{(-1)^{N}e^{[(\tau - \dario + N\omega)(2N\zeta(\omega))]}}{(-1)^{N}e^{[(\tau + \dario + N\omega)(2N\zeta(\omega))]}} 
\end{equation*}
which becomes:
\begin{equation}
\phi(\tau+N T_\tau)  = \phi(\tau) e^{-4 N \dario \zeta(T_\tau / 2)}
\end{equation}
valid also in the case of $g_3 < 0$ as can be shown repeating the above computation for $T_\tau = -i\omega$ and using the identity $\wp(z,g_2,g_3) = -\wp(iz,g_2,-g_3)$.
The following now holds:
$$
e^{i(v_mNT_\tau-\Delta\theta)} = e^{4N((T_\tau / 2)\zeta(\dario)- \dario \zeta(T_\tau / 2))}
$$
and, for $N=1$:
$$
\Delta\theta = v_m T_\tau - 4 \mbox{Im}\left[T_\tau / 2 \zeta(\dario)- \dario \zeta(T_\tau / 2)\right]
$$
hence the condition for periodic motion:
$$
v_m T_\tau - 4 \mbox{Im}\left[T_\tau / 2 \zeta(\dario)- \dario \zeta(T_\tau / 2)\right] = 2 q \pi 
$$
where $q = M / N \in \mathbb Q$ is rational. The trajectory plotted in Figure \ref{fig:traj} was
found iteratively by finding $v_p$ so that in the above equation $q=1/10$.

\section{Conclusions}
An exact, explicit, closed form, solution of the constant radial acceleration problem can be written relating the state to a pseudo-time. The solution is elegantly given, in all cases, by an expression involving Weierstrass elliptic and related functions. Just like in the Keplerian mechanics, a radial Kepler equation must then be solved to recover the time dependance.
Such a solution adds to the list of interesting problem of classical mechanics that can be solved
by the use of Weierstrass elliptic and related functions and provides a new useful tool for aerospace
engineers and physicists who deal with the application of this dynamics.

\section*{References}

\bibliographystyle{aiaa}
\bibliography{biblio_radial}

\end{document}